\documentstyle[12pt]{article}
\begin{document}

\title{The non-deterministic quantum logic
operation and teleportation of a vacuum and single photon
superposition state via parametric amplifiers}
\author{XuBo Zou, K. Pahlke and W. Mathis\\
Electromagnetic Theory Group at THT,\\
 Department of Electrical
Engineering, \\
University of Hannover, Germany}
\date{}

\maketitle

\begin{abstract}
{\normalsize We firstly present an all-optical scheme to implement
the non-deterministic quantum logic operation of Knill, Laflamme
and Milburn (Nature, 409, 46-52(2001)). In our scheme, squeezed
vacuum state is acted as auxiliary state instead of single photon
resources. Then we demonstrate that same setup can be used to
teleport a superposition of vacuum and single photon state of the
form $\alpha|0>+\beta|1>$ and a superposition of vacuum and single
polarized photon state of the form $\alpha|0>+\beta|H>+\gamma|V>$.
}

\end{abstract}
\section * {I. INTRODUCTION}
The development of quantum algorithms shows impressively, that a
quantum computer can provide an enormous speed up compared to
classical computers \cite{shor}\cite{gro}. This principle
theoretical result motivated an intensive research in the field of
quantum information processing. Several physical systems were
suggested to implement the concept of quantum computing: cavity
QED systems\cite{tc} and trapped ion systems\cite{cd},quantum dot
systems \cite{loss} and Josephson-junction device systems
\cite{mak}. The realization of a linear optical quantum computer
is particularly appealing because of the robust nature of quantum
states of light against the decoherence. Recently the
implementation of a probabilistic quantum logic gate was proposed,
which is based on linear optical elements\cite{klm}. A
non-deterministic gate, which is known as the nonlinear sign-gate
(NLS), was proposed in this paper. This gate transforms the
quantum states, which are labeled by the photon occupation number,
according to
\begin{eqnarray}
\alpha|0\rangle+\beta|1\rangle+\gamma|2\rangle\rightarrow\alpha|0\rangle+\beta|1\rangle-\gamma|2\rangle
\,.\label{1}
\end{eqnarray}
The probability of success is $p_c=1/4$. Although the scheme in
Ref \cite{klm} contains only linear optical elements, the optical
network is complex and would present major stability and mode
matching problems in their construction. Several less complicated
schemes with linear optical elements were presented
\cite{ra,ru,zou} to implement the NLS-gate with a slightly lower
probability of success. In addition, if the information is stored
in the polarization of photons, a probabilistic quantum logic gate
(CNOT-gate and controlled-phase gate) can also be implemented
\cite{bbo,pi}. But these schemes require single photon or
entangled photon states as a auxiliary resource. The requirement
of single photon resources is a difficult constraint to satisfy
with available technology, while currently triggered single photon
sources operate by means of fluorescence from a single molecule
\cite{cbb} or a single quantum dot \cite{cs,pm}, which have
terrible spatial properties, and do not meet the requirement for
linear quantum information processing. As described in
Ref\cite{klm}, simple photon resource can be generated using weak
squeezing and particle detectors. A non-degenerate squeezer is
applied to the vacuum state, the output consists of the a
superposition of the states with identical numbers of the bosons
in the two modes. Thus a single photon state can be produced in
one of the modes conditioned on the the single photon detection of
another mode. In this case, a all-optical non-deterministic
quantum logic operations can be implemented by using the four
photon modes and three photon detectors with lower probability of
success $p_sp_c$. Here $p_s<0.25$ is probability of generation of
single-photon state by using non-degenerate squeezer. More
recently an experiment is implemented for realizing a control
conditional phase gate using parametric down-conversion,
which is based on the post-selective method\cite{ste}.\\
Quantum teleportation, invented by Bennett et al\cite{bb}, is one
of the essential primitive of quantum communication, which
transmit a quantum state from one observer to another observer
through dual EPR state\cite{epr} and classical communication.
Since the first successful realization of teleportation was
reported\cite{a1}, the teleportation concept is extended to the
continuous variable\cite{a3}.  A number of schemes have been
presented for teleportation of discrete
system[\cite{bh}-\cite{hwl}]. In Ref\cite{hwl}, schemes have been
proposed to teleport a running wave superposition of zero and
one-photon field state , in which entangled state channel is
prepared through a single photon state incident a symmetric beam
splitter. For continuous variable teleportation\cite{a3}, the
first experimental realization is to teleport a coherent state by
means of the parametric amplifier and quadrature-component
measurement\cite{a2}. It has also been realized that the
number-difference, phase-sum teleportation protocols can be
realized by using parametric amplifier and the measurement of
phase\cite{a6}. Further, Clausen et al proposed a variation on the
scheme of Milburn and Braunstein avoiding problem associated with
the measurement of phase and demonstrate the Fock state can be
teleported with a fidelity equal to unity\cite{a8}. In
Ref\cite{a10}, continuous variable teleportation of single photon
state was investigated using normal teleportation
protocols\cite{a3,a2}. In this case, the output state is different
from the single photon state due to the nonmaximal entanglement
channel.\\
The main purpose of this paper is twofold. First, we firstly
present an all-optical scheme to implement the non-deterministic
quantum logic operation(1) using two non-degenerate squeezers and
a two-photon coincidence detection. Our scheme does not need
single photon resource as auxiliary resource and require three
photon modes. Second, we show that the same setup can be used to
teleport a superposition of vacuum and single photon state of the
form $\alpha|0>+\beta|1>$ and a superposition of vacuum state and
single polarized photon state of the form
$\alpha|0>+\beta|H>+\gamma|V>$, which illustrate the quantum
teleportation with three-state system, namely, the basis states
are $\{|0>,|H>,|V>\}$.
\section * {II. THE NON-DETERMINISTIC QUANTUM LOGIC OPERATION}
Our scheme is based on the teleportation protocols\cite{bb}. To
use teleportation schemes for universal quantum computation was
first suggested by Gottesman and Chuang \cite{chu}. Since the
scheme of the Knill et al\cite{klm}, several scheme has been
proposed for linear optical quantum
computation\cite{ra}-\cite{pi}. But these schemes require
entangled photon states or single photon state as an auxiliary
resource. Recently an experiment is implemented for realizing a
control conditional phase gate using parametric
down-conversion\cite{ste}. However, the scheme is based on the
post-selective method, which limit application of the scheme. Now
we will describe the realization of the non-deterministic quantum
logic operation(1), in which an squeezed vacuum state is auxiliary
photon state. The experimental setup is depicted in Fig.1, which
require two non-degenerate squeezers and a two-photon coincidence
detection. The input quantum state of the mode 1 is in the form
\begin{eqnarray}
\Psi_{in}=\alpha|0\rangle_1+\beta|1\rangle_1+\gamma|2\rangle_1\,.
\label{2}
\end{eqnarray}
An ancilla quantum state in the vacuum state
$\Psi_{ancilla}=|0\rangle_2|0\rangle_3$ is required. After mode 2
and 3 passing through the first parametric amplifier, whose
transformation is given by
$$
S_1=\exp[\theta_1(a_2^{\dagger}a_3^{\dagger}-a_2a_3)]) \eqno{(3)}
$$
where $a_2$ and $a_3$ are bose annihilation operators. This
expression can be rewritten in a disentangled form
$$
U_3=\exp(\sqrt{\gamma_1}a_2^{\dagger}a_3^{\dagger})(1-\gamma_1)^{(a_2^{\dagger}a_2+a_3^{\dagger}a_3+1)/2}\exp(-\sqrt{\gamma_1}a_2a_3)
\eqno{(4)}
$$
where $\gamma_1=\tanh^2\theta_1$. Thus the output state of mode 2
and 3 is
$$
\Psi^{\prime}_{ancilla}=
\sqrt{1-\gamma_1}\sum_{n=0}^{\infty}\gamma_1^{n/2}|n\rangle_{2}|n\rangle_{3}\,.
\eqno{(5)}
$$
The coefficients $\gamma_1$ will be determined later. The field
state $\Psi_{in}\Psi^{\prime}_{ancilla}$ will directly be used for
the NLS-gate implementation. One of the the two output modes of
the first parametric amplifier is then used as one of the input
modes of the second parametric amplifier, whose transformation is
given by
$$
S_2=\exp[\theta_2(a_1^{\dagger}a_2^{\dagger}-a_1a_2)]) \eqno{(6)}
$$
and the mode whose quantum state $\Psi_{in}$ is desired to be
operated is the other input mode of the second parametric
amplifier. After mode 1 and 2 passing through the the second
parametric amplifier, the state of the system become
$$
\Psi_{out}=\sqrt{(1-\gamma_1)(1-\gamma_2)}[\sqrt{\gamma_2}\alpha|0>_3+\sqrt{\gamma_1}(1-2\gamma_2)\beta|1>_3
$$
$$
+\gamma_1\sqrt{\gamma_2}(3\gamma_2-2)\gamma|2>_3]|1>_1|1>_2
+\Psi_{other} \eqno{(7)}
$$
The terms $\Psi_{other}$ of this quantum state don't contribute to
the events, which correspond to a 2-photon coincidence (one photon
in each of the beams). The output state of the system, conditioned
on the two-fold coincidence detection, is projected into
$$
\Psi_{out}=\sqrt{(1-\gamma_1)(1-\gamma_2)}[\sqrt{\gamma_2}\alpha|0>_3+\sqrt{\gamma_1}(1-2\gamma_2)\beta|1>_3
+\gamma_1\sqrt{\gamma_2}(3\gamma_2-2)\gamma|2>_3] \eqno{(8)}
$$
If we choose the parameter $\gamma_1$ and $\gamma_2$ to satisfy
$$
\sqrt{\gamma_2}=\sqrt{\gamma_1}(1-2\gamma_2)=-\gamma_1\sqrt{\gamma_2}(3\gamma_2-2)
\eqno{(9)}
$$
i.e. $\gamma_1=(21-7\sqrt{2})/(9+4\sqrt{2})\approx0.757$ and
$\gamma_2=(3-\sqrt{2})/7\approx0.226$,  the output quantum state
$\Psi_{out}=\alpha|0\rangle+\beta|1\rangle-\gamma|2\rangle$ is
generated from the input quantum state $\Psi_{in}$. This is the
transformation property(1) of the NLS-gate. The probability of
success of our scheme will be approximate $4.25\%$. In scheme
proposed in Ref\cite{klm} and \cite{ra,ru}, if the single-photon
resource is generated by using weak squeezing and particle
detectors, a all-optical NLS gate can be implemented with the
maximally probability of success $6.25\%$ and $5.6\%$,
respectively, which require four field modes and three photon
detectors. Our scheme is comparatively simple, which only require
two parametric amplifiers, three field modes and a two-photon
coincidence detection. Since $\gamma_1\approx0.757$, this
demonstrate the scheme work in strong coupling regime. The
currently available parametric amplifier can produce strongly
squeezed light. Aytur and Kumar\cite{kum} have reported on a
parametric gain $\gamma_1\approx0.9$ in the Eq.(5). Another
requirement of the scheme is two-photon coincidence detection
which has been used for many polarization photon quantum
information processing experiments in the weak coupling
regimes\cite{a1,a12}. In the strong coupling regimes, it is
required that these detectors should be capable of distinguishing
different photon number. This requirement is met in proposed
teleportation scheme involving photon number
measurement\cite{klm,ra,ru,zou,a6,a8,a13}.
\section * {III. THE TELEPORTATION OF A VACUUM AND SINGLE PHOTON SUPERPOSITION SATATE}
In Ref\cite{hwl}, schemes have been proposed to teleport a running
wave superposition of zero and one-photon field state , in which
entangled state channel is prepared through a single photon state
incident a symmetric beam splitter. In Ref\cite{a1}, teleportation
of the polarization state of single photon has been observed. In
Ref\cite{a10}, it was shown that single photon polarization state
was not teleported with a fidelity equal to unity by using
parameters and quadrature phase measurement. In this section, we
will demonstrate that setup shown in Fig.1 can also be used to
teleport a superposition state of zero and single photon
$\alpha|0>+\beta|1>$ and a superposition of vacuum state and
single polarized photon state of the form
$\alpha|0>+\beta|H>+\gamma|V>$.\\
Firstly we consider the teleportation of a superposition state of
zero and single photon $\alpha|0>+\beta|1>$. We set $\gamma=0$ in
Eq.(2) and repeat the same process suggested in Sec.II. The output
state of the system, conditioned on the two-fold coincidence
detection, is projected into
$$
\Psi_{out}=\sqrt{(1-\gamma_1)(1-\gamma_2)}[\sqrt{\gamma_2}\alpha|0>_3+\sqrt{\gamma_1}(1-2\gamma_2)\beta|1>_3
] \eqno{(10)}
$$
If we choose the parameter $\gamma_1$ and $\gamma_2$ to satisfy
$$
\sqrt{\gamma_2}=\sqrt{\gamma_1}(1-2\gamma_2) \eqno{(11)}
$$
i.e. $\gamma_1=\gamma_2/(1-2\gamma_2)^2$, we can obtain the
expected teleported state
$\Psi_{out}=\alpha|0\rangle+\beta|1\rangle$ with probability of
success $2(1-\gamma_1)(1-\gamma_2)\gamma_2$. If $\gamma_2<<1$,
then $\gamma_1<<1$. This means that two parametric amplifier can
simultaneously work in the weak coupling regimes, which can be
attained by the present experimental technology. In this weak
regime, a number of experiments have been implemented for quantum
information processing\cite{a1,a12}. Another requirement of the
scheme is two-photon coincidence detection. In the weak coupling
regime, two-photon coincidence detection has been used for
teleportation of single polarization photons in Ref\cite{a1}. This
demonstrate that the present scheme can be implemented with
present technology.\\
So far, we have only considered teleportation of optical fields in
zero and one photon-number superposition. In addition, one can
introduce the polarization of the field, which may be used to
encode information. It is interesting to teleport superposition
state of the vacuum and single polarized photon of the form
$\alpha|0>+\beta|H>+\gamma|V>$, which illustrate the quantum
teleportation with three-state system, namely, the basis states
are $\{|0>,|H>,|V>\}$. In this case, the input quantum state of
the path 1 is in the form
$$
\Phi_{in}=\alpha|0>_1+\beta|H>_1+\gamma|V>_1\,. \eqno{(12)}
$$
Now we consider pulsed type-II frequency-degenerate parametric
down conversion(PDC), which can be described by the interaction
$$
H=i\gamma(a_{H2}^{\dagger}a_{V3}^{\dagger}+a_{V2}^{\dagger}a_{H3}^{\dagger}-a_{H2}a_{V3}-a_{V2}a_{H3})
\eqno{(13)}
$$
where $a_{H2}^{\dagger}$ is the creation operator for a photon
with polarization H propagating along path 2 etc. This is the
familiar Hamiltonian for the creation of polarization entangled
photon pairs, which has been the basis for many experiments in
quantum information\cite{a1,a12}. For initial vacuum state, one
can show that this interaction leads to
$$
\Phi^{\prime}_{ancilla}=
({1-\gamma_1})\sum_{n=0}^{\infty}\sqrt{n+1}\gamma_1^{n/2}\Psi^{n}
\eqno{(14)}
$$
where $\gamma_1=\tanh^2(\gamma t)$ and
$$
\Psi^{n}=\frac{1}{\sqrt{n+1}n!}(a_{H2}^{\dagger}a_{V3}^{\dagger}+a_{V2}^{\dagger}a_{H3}^{\dagger})^n|0>
\eqno{(15)}
$$ The output 2 of the first Type-II PDC is then
used as one of the input of the second Type-II PDC and the
$\Phi_{in}$ is desired to be teleported is the other input of the
second Type-II PDC. After passing through the the second Type-II
PDC, the state of the total system become
$$
\Phi_{out}={(1-\gamma_1)(1-\gamma_2)}\sqrt{\gamma_2}[\sqrt{\gamma_2}\alpha|0>_3+\sqrt{\gamma_1}(1-2\gamma_2)\beta|H>_3
$$
$$
+\sqrt{\gamma_1}(1-2\gamma_2)\gamma|V>_3]|HV>_1|HV>_2
+\Psi_{other} \eqno{(16)}
$$
where $\gamma_2$ is squeezed parameter of second Type-II PDC.  The
terms $\Psi_{other}$ of this quantum state don't contribute to the
events, which correspond to a four-photon coincidence
$|HV>_1|HV>_2$. The output state of the system, conditioned on the
two-fold coincidence detection, is projected into
$$
\Phi_{out}=\sqrt{(1-\gamma_1)(1-\gamma_2)}[\sqrt{\gamma_2}\alpha|0>_3+\sqrt{\gamma_1}(1-2\gamma_2)(\beta|H>_3+\gamma|V>_3)
] \eqno{(17)}
$$
If we choose the parameter $\gamma_1$ and $\gamma_2$ to satisfy
equation(11),  we can obtain the expected teleported state
$\Phi_{out}=\alpha|0>+\beta|H>+\gamma|V>$ with probability of
success $3(1-\gamma_1)^2(1-\gamma_2)^2\gamma_2^2$. For the
experimental realization of the present scheme, a four-photon
coincidence detection is needed. In the weak coupling regime,
four-photon coincidence detection has been used for generation of
four-photon GHZ state\cite{a12}. Fortunately our scheme can work
in weak coupling,($\gamma_2<<1$ and $\gamma_1<<1$), so that our
scheme can be implemented with the present experimental
technology.
\section * {IV. CONCLUSION}
In summary, we have presented scheme for implementing a
all-optical non-deterministic NLS gate and teleporting a
superposition state of zero and single photon $\alpha|0>+\beta|1>$
and transmission of superposition of vacuum state and a polarized
photon state of the form $\alpha|0>+\beta|H>+\gamma|V>$, which
require two non-degenerate squeezers and a multi-photon
coincidence detection. Our scheme does not need single photon
resource as auxiliary resources and only require three photon
modes. However the price of avoiding single photon resource is a
low success probability of NLS gate. We notice that the
experimental setup of the present scheme has been illustrated in
famous experiments by zou et al\cite{zzz}, where parametric
down-conversion has been used to show an interference experiment.
This has been accomplished by aligning the idler beam of the first
parametric amplifier with the idler beam of the second parametric
amplifier. For the implementation of a all-optical
non-deterministic NLS gate, one requirement of our scheme is a
strong coupling amplifier. The currently available parametric
amplifier can produce strongly squeezed light\cite{kum}, which can
satisfy the requirement of the present scheme. Thus, the main
difficulty of our scheme in respect to an experimental
demonstration consists in the requirement on the sensitivity of
the detectors. The present scheme require two-photon coincidence
detection. In the weak coupling regimes, two-photon coincidence
detection has been used for many polarization photon quantum
information processing experiments\cite{a1}. In the strong
coupling regimes, it is required that these detectors should be
capable of distinguishing different photon number. Recently,
experimental techniques for photon detection made tremendous
progress. A photon detector based on visible light photon counter
can distinguish between a single photon incidence and two photon
incidence. A high quantum efficiency with a good time resolution
and a low bit-error rate was reported \cite{yyy}. However, in our
scheme, we need a photon detector which should be capable of
distinguishing different photon number. To fulfill the
requirements of the presented scheme will be experimentally
challenging. For the realization of teleportation of a vacuum and
single photon superposition state ,the present scheme can operate
in weak regime and require two-photon or four-photon coincidence
detection. These requirement can be realized experimentally, which
means that our scheme can be realized with the present
experimental technology. The result imply that unconditional
continuous variable teleportation of single photon polarization
could be considered an alternative to the postselected
scheme\cite{a1} using entangled photon state.

\begin{flushleft}

{\Large \bf Figure Captions}

\vspace{\baselineskip}

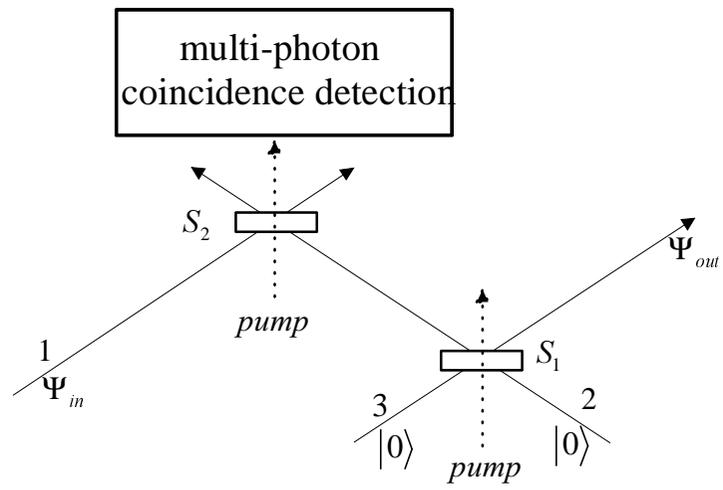
{\bf Figure1 .} The schematic is shown to implement the
non-deterministic quantum logic operation, which has been
explained in the paper.

\end{flushleft}

\end{document}